\documentclass[12pt]{article}
\usepackage{epsfig}
\title{A gauge theory of the hamiltonian
reduction for the rational
Calogero - Moser system.}
\author{Cezary Gonera\thanks{supported by KBN grant 2 P03B 134 16},
 Piotr Kosi\'nski$^*$, Pawe{\l} Ma\'slanka$^*$ \\
Department of Theoretical Physics II \\
University of {\L}\'od\'z \\
Pomorska 149/153, 90 - 236 {\L}\'od\'z/Poland.}
\date{}

\begin{document}
\maketitle
\begin{abstract}
A gauge theory equivalent to the hamiltonian
reduction scheme for rational Calogero - Moser
model is presented.
\end{abstract}

\section{Introduction}
Hamiltonian reduction 
\cite{b1}, \cite{b2} is one of the most powerful methods of constructing integrable models.
In particular, the celebrated Calogero - Moser system can be described within this framework, both in degenarate \cite{b3}, \cite{b4} as 
well as in general elliptic case \cite{b5}.\\
The general reduction scheme can be described as follows. There is a symplectic manifold $(M,\omega)$\ on which a Lie group G acts in a symplectic way.
It is assumed that this action is  strongly hamiltonian.
One selects a G - invariant function H on M as a hamiltonian of the dynamical system to be reduced.
Clearly, H has vanishing Poisson brackets with all hamiltonians generating the action of G. The reduced phase space is defined as follows. Let $g $\ be the Lie algebra of G; for any $\xi \in g$\  let  $h_{\xi}$\ denote the corresponding hamiltonian. Obviously, $h_{\xi}$\ is a linear function of $\xi$\ , so that $h_{\xi}(p),$\ $p \in M$\ defines an element of $g^{*};$\ 
the mapping $\mu : M \rightarrow g^{*},$\ $<\mu(p),\xi> = h_{\xi}(p)$\ is called the momentum map.  In order to define the reduced phase space one selects an element $\alpha \in  g^{*}$\ and consider the subset $P_{\alpha} \subset M$\ corresponding to $\mu(p) = \alpha.$\
Under suitable assumptions $P_{\alpha}$\ is a submanifold of M. However, $P_{\alpha}$\ is in general not sympletic: $\omega \mid_{P_{\alpha}}$\ is
     degenerate. Fortunately, this degeneracy can be easily characterized: let $G_{\alpha} \subset G$\ be a stability subgroup of $\alpha$\ under coadjoint action.
It appears then that $Q_{\alpha} = P_{\alpha}/G_{\alpha}$\ (which, again, under some suitable assumptions is a manifold ) is symplectic, i. e. $\omega\mid_{Q_{\alpha}}$\ is nondegenerate.
 Now, with H being G - invariant the momentum map is 
 a constant of motion so that $P_{\alpha}$\ is invariant under dynamics generated by H. The essence
of the method is that the trajectories on $P_{\alpha}$\ when projected on $Q_{\alpha}$\ are hamiltonian, the hamiltonian function being $H \mid_{Q_{\alpha}}.$\
\\
\\
\psfig{figure=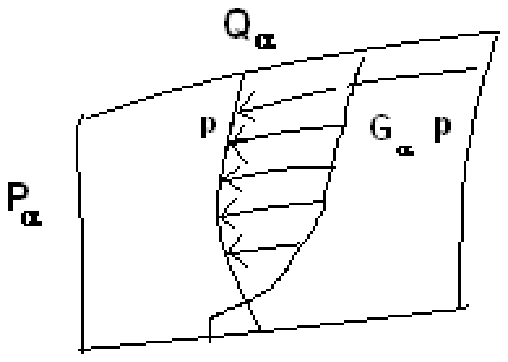,height=5cm}
\\
\\
One can now pose the question what is the lagrangian theory behind this scheme. To answer this question let us note that
 we are eventually looking for is the dynamics in terms of canonical variables on $Q_{\alpha}$\ . It is obvious that any function of these variables can be regarded as $G_{\alpha}$\ - invariant function on $P_{\alpha}$\ ( any $G_{\alpha}$\ invariant function on $P_{\alpha}$\ is constant on $G_{\alpha}$\ orbits so it is a function on $Q_{\alpha}$\ ).
Therefore, a given reduced trajectory on $Q_{\alpha}$\ corresponds to an infinite set of curves on $P_{\alpha},$\ all of them being related to each other by the time - dependent action of $G_{\alpha}.$\
\\
\\
\psfig{figure=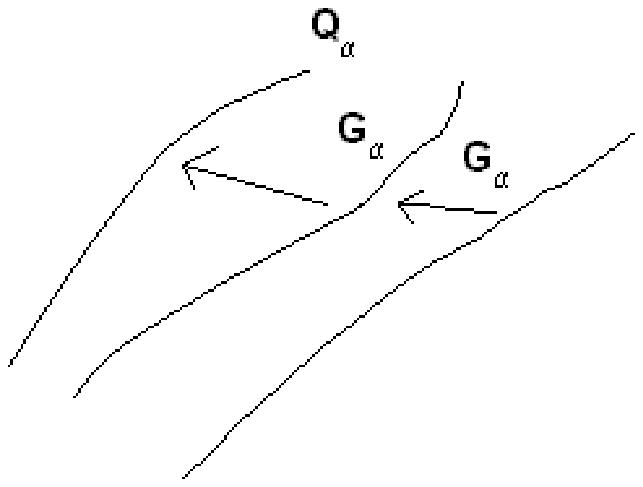,height=5cm}
\\
\\
This suggests strongly that the corresponding lagrangian dynamics possess gauge symmetry related 
to $G_{\alpha}.$\ \\
It seems that the above reasoning is fairly general, i.e. the hamiltonian reduction can be described within the framework of lagrangian gauge
theory. The details of this general scheme will be published  elsewhere.
 Here we consider a particular example of ( degenerate) Calogero - Moser systems \cite{b6},
 \cite{b7}. An explicit construction of relevant gauge theory has been given by 
Polychronakos \cite{b8}.
In order to obtain a lagrangian theory he made first the trick consisting essentially in taking $G_{\alpha} = G.$\ This is  achieved by viewing the value $\alpha$\ of momentum map as a dynamical variable transforming under G according to coadjoint representation. The advantage is that the constraint concerning the value of momentum map becomes now G - invariant, $G_{\alpha} = G.$\ On the other hand
the new dynamical variables do not appear in hamiltonian, so they are frozen on any particular trajectory. The reduced theory appears then to be the gauge theory with the gauge group G while the reduced dynamics corresponds to the temporal gauge. \\
The aim of the present note is to construct the lagrangian dynamics for Calogero - Moser model without enlarging the set of dynamical variables. It is not surprising that the resulting lagrangian theory is a gauge theory with the gauge group G broken down explicilty to $G_{\alpha}.$\ \\
 \section{Calogero model as a gauge theory.}
(i) \underline{ Hamiltonian reduction}\\
We start with a brief description of general reduction method mentioned above as applied to rational Calogero - Moser model \cite{b3}.\\
As an unreduced phase space T one takes the set of pairs of traceless hermitian nxn matrices A and B.
\begin{equation}
T=\{(A,B); A=A^{+}, B=B^{+}, TrA = 0 = TrB\}    \\
\end{equation}
 The symplectic form on T is given by
\begin{equation}
\omega = Tr (dB \wedge dA)\\
\end{equation}
and implies following Poisson brackets
\begin{equation}
\{A_{ij}, B_{mn}\} = \delta_{in} \delta_{jm} - \frac{1}{N}\; \delta_{ij} \delta_{mn}\\
\end{equation}
There is a natural symplectic action of SU(N) group on T
\begin{equation}
SU(N)\ni U : (A,B) \rightarrow (UAU^{+}, UBU^{+})\\ 
\end{equation}
which generates a hamiltonian vector fields $\zeta$\ on T 
\begin{equation}
\zeta (A,B) = (\{A,H_{\widehat{\zeta}}\}, \{B,H_{\widehat{\zeta}}\})\\
\end{equation}
with hamiltonians
\begin{equation}
H_{\widehat{\zeta}} (A,B) = Tr \; (i [A,B] \widehat{\zeta}),\;\;\; \widehat{\zeta} \in sU(N) \\
\end{equation}
Hence, the momentum map reads 
\begin{equation}
\Phi (A,B) = i [A,B] \\                                   
\end{equation}
Dynamics on our unreduced phase space is defined by SU(N)-invariant hamiltonian
\begin{equation}
H(A,B) = \frac{1}{2} TrB^{2}
\end{equation}
The element $\alpha$\ of sU(N) algebra and the reduced phase space $Q_{\alpha}$\ constructed along lines sketched in the introduction read
\begin{eqnarray}
&&\alpha = ig(v^{+} \otimes v - I); \;\;\;\; v=(1,1,...,1) \\
\nonumber \\
&&Q_{\alpha} = P_{\alpha}/G_{\alpha} \nonumber \\
\nonumber \\
&&P_{\alpha} = \{(A,B) \in T \; ; [A,B] = \alpha\} \nonumber \\
\nonumber \\
&&G_{\alpha} = \{ C \in SU(N) \;\; C \alpha C^{+} = \alpha \} \nonumber
\end{eqnarray}
In other words, $Q_{\alpha}$\ consists of pairs of matrices (Q,L) which solve momentum map equation  
$ \Phi(A,B) = \alpha $\
and cannot be related by $G_{\alpha}$\ action. \\
One can show that these matrices can be parametrized in the following nice way 
\begin{eqnarray}
Q(q)=diag(q_{1}...q_{N})  \nonumber \\
\nonumber \\
(L(q,p))_{ij}=p_{j}\delta_{ij}+ig(1-\delta_{ij}) \frac{1}{q_{i}-q_{j}}; \label{w10} \;
\end{eqnarray}

here $L(p,q)$\ is nothing but the Lax matrix of rational Calogero - Moser system.\\
In such a way the unreduced dynamics given by $H(A,B)=\frac{1}{2} TrB^{2}$\ with constraint $\Phi(A,B)=\alpha$\
when solved "modulo" the transformations from stability group $G_{\alpha}$\ provides the rational C-M model.\\
(ii) \underline{Lagrangian gauge theory} \\
Now, to discuss the equivalence of rational C-M model to some gauge theory let us consider the SU(N) gauge theory (we assume that we are in the center - of - mass system) broken explicitly
by linear term. The "matter" transforms according to the adjoint representation of SU(N). The simplest lagrangian reads \\
\begin{eqnarray}
L(q,\dot{q}) = \frac{1}{2}(\dot{q}_{\alpha}+ f_{\alpha \beta \gamma}q_{\beta}A_{\gamma})^{2} - v_{\alpha}A_{\alpha}\;\;, \label{w11}
\end{eqnarray}
where $f_{\alpha \beta \gamma}$\ are SU(N) structure constants, $v_{\alpha}$\ is a fixed vector in adjoint representation of SU(N) and $A_{\alpha}$\ is a gauge field. L
is invariant ( up to a total derivative ) under the stability subgroup $G_{v}\subset  SU(N) $\ of  $v_{\alpha}:$\ \\
\begin{eqnarray}
&&q_{\alpha}\rightarrow q'_{\alpha} = D_{\alpha \beta}(t)_{q_{\beta}} \nonumber\\
&&A_{\alpha}\rightarrow A'_{\alpha} = D_{\alpha \beta}(t)A_{\beta} - \varrho_{\alpha}(t), \; f_{\alpha \beta \gamma}\varrho_{\gamma}=\dot{D}_{\alpha \varrho}(t)D^{-1}_{\varrho \beta}(t) \label{w12}\; \\
&&D_{\alpha \beta}(t)v_{\beta}=v_{\alpha}\nonumber
\end{eqnarray} 
The corresponding equations of motion read 
\begin{eqnarray}
f_{\alpha \beta \gamma}q_{\beta}(\dot{q}_{\alpha}+f_{\alpha \zeta \sigma}q_{\zeta}A_{\sigma})-v_{\gamma}=0  \label{w13}\; \\
\frac{d}{dt}(\dot{q}_{\alpha}+f_{\alpha \beta \gamma}q_{\beta}A_{\gamma})-
f_{\alpha \beta \gamma}A_{\beta}(\dot{q}_{\gamma}+f_{\gamma \zeta \sigma}q_{\zeta}A_{\sigma})=0  \label{w14}\; 
\end{eqnarray}
Let us pass to the hamiltonian formalism.
The canonical momenta are
\begin{eqnarray}
&&p_{\alpha}\equiv\frac{\partial L}{\partial q_{\alpha}}=\dot{q}_{\alpha}+f_{\alpha \beta \gamma}q_{\beta}A_{\gamma} \label{w15}\; \\
&&\pi_{\alpha}\equiv\frac{\partial L}{\partial \dot{A}_{\alpha}}=0 \nonumber
\end{eqnarray}
The theory is constrained, $\pi_{\alpha}\approx 0 $\ being the primary constraints.
The hamiltonian is constructed according to the standard rules \cite{b9}.
\begin{eqnarray}
H=p_{\alpha}q_{\alpha}-L=\frac{1}{2}p_{\alpha}^{2}-f_{\alpha \beta \gamma}p_{\alpha}q_{\beta}A_{\gamma}+
A_{\alpha}v_{\alpha} \label{w16}\;
\end{eqnarray}
and yields the following canonical equations
\begin{eqnarray}
\dot{q}_{\alpha}&=&p_{\alpha}-f_{\alpha \beta \gamma}q_{\beta}A_{\gamma} \label{w17}\ \\
\dot{A}_{\alpha}&=&u_{\alpha} \label{w18}\ \\
\dot{p}_{\alpha}&=&f_{\alpha \beta \gamma}A_{\beta}p_{\gamma} \label{w19} \\
\dot{\pi}_{\alpha}&=&f_{\alpha \beta \gamma}p_{\beta}q_{\gamma}- v_{\alpha} \label{w20}
\end{eqnarray}
resulting from $\tilde{H}\equiv H+u_{\alpha}\pi_{\alpha}, \; u_{\alpha}$\ being the Lagrange
 multipliers.\\
Following the standard procedure \cite{b9} we obtain the secondary constraints. First, $\dot{\pi}_{\alpha}\approx 0 $\ 
gives, together with (\ref{w20}),
\begin{eqnarray}
\chi_{\alpha}\approx f_{\alpha \beta \gamma}p_{\beta}q_{\gamma}-v_{\alpha}=0 \label{w21}
\end{eqnarray}
which is nothing but the momentum map condition.
Taking again a time derivative of 
(\ref{w21}) and using Jacobi identity we arrive at 
\begin{eqnarray}
\tilde{\chi}_{\alpha}\equiv f_{\alpha \beta \gamma}v_{\beta}A_{\gamma}\approx 0 \label{w22}
\end{eqnarray}
This completes the list of constraints since $\dot{\chi}_{\alpha}\approx0$\ yields
\begin{eqnarray}
f_{\alpha \beta \gamma}v_{\beta}u_{\gamma}\approx0 \label{w23}
\end{eqnarray}
which is a constraint for Lagrange multipliers.\\
The full set of constraints reads 
\begin{eqnarray}
\pi_{\alpha}\approx 0 \label{w24} \; - \;\hbox{the primary  constraints}
\end{eqnarray}
\begin{equation}
\left.
\begin{array}{ccl}
\chi_{\alpha}&\equiv& f_{\alpha \beta \gamma}p_{\beta}q_{\gamma} - v_{\alpha}=0  \\
\tilde{\chi}_{\alpha}&\equiv& f_{\alpha \beta \gamma}u_{\beta}A_{\gamma}=0 
\end{array}
\right\}\;\;\;-\;\hbox{the secondary constraints} \label{w25}
\end{equation}
The nontrivial Poisson brackets are 
\begin{eqnarray}
\{\pi_{\alpha},\;\tilde{\chi}_{\beta}\}&=&\;-f_{\alpha \beta \gamma}v_{\gamma} \label{w27} \\
\{\chi_{\alpha},\;\chi_{\beta}\}&=&\;-f_{\alpha \beta \gamma}(\chi_{\alpha}+v_{\gamma}) \nonumber
\end{eqnarray}
The next step is to identify the first- and second-class constraints. Let $a_{\underline{\alpha}}$\ be the generators of
$G_{v},\; (a_{\underline{\alpha}})_{\varrho \sigma} \equiv a_{\underline{\alpha} \beta}f_{\beta \varrho \sigma}$, 
while $b_{\underline{\underline{\alpha}}}\;((b_{\underline{\underline{\alpha}}})_{\varrho \sigma}\equiv
b_{\underline{\underline{\alpha}}\beta}f_{\beta \varrho \sigma})$\ - the remaining ones;\\
Obviously
\begin{eqnarray}
a_{\underline{\alpha}\beta}f_{\beta \varrho \sigma}v_{\sigma}=0 \label{w28}
\end{eqnarray}
Let us define
\begin{eqnarray}
\pi_{\underline{\alpha}}&\equiv& a_{\underline{\alpha}\beta}\pi_{\beta}, \; 
\pi_{\underline{\underline{\alpha}}} \equiv b_{\underline{\underline{\alpha}} \beta}\pi_{\beta} \nonumber \\
\chi_{\underline{\alpha}}&\equiv& a_{\underline{\alpha} \beta}\chi_{\beta}, \; 
\chi_{\underline{\underline{\alpha}}}\equiv b_{\underline{\underline{\alpha}}\beta}\chi_{\beta} \label{w29} \\
\tilde{\chi}_{\underline{\alpha}}&\equiv& a_{\underline{\alpha}\beta}\tilde{\chi}_{\beta}, \;
\tilde{\chi}_{\underline{\underline{\alpha}}} \equiv b_{\underline{\underline{\alpha}}\beta}
\tilde{\chi}_{\beta} \nonumber
\end{eqnarray}
It follows then from (\ref{w28}) that $ \tilde{\chi}_{\underline{\alpha}}=0$\ as an identity while
\begin{eqnarray}
\{\pi_{\underline{\alpha}},\; \bullet \; \} \approx 0 \label{w30} \\
\{\chi_{\underline{\alpha}}, \; \bullet \;\} \approx 0 \nonumber
\end{eqnarray}
which means that $\pi_{\underline{\alpha}}$\ and $\chi_{\underline{\alpha}}$\ are first class constraints. In 
order to deal with the remaining ones let us denote by $(a_ 
{\beta{\underline{\gamma}}}, \;\tilde{b}_{\beta \underline{\gamma}})$\ 
the matrix inverse to $ (a_{\underline{\alpha}\beta},\;  b_{\underline{\underline{\alpha}}\beta})^{T} 
 $\

\begin{eqnarray}
&a_{\underline{\alpha}\beta}\tilde{a}_{\beta \underline{\gamma}}&= \delta_{\underline{\alpha} \underline{\gamma}},\;\;\;
a_{\underline{\alpha}\beta}\tilde{b}_{\beta \underline{\underline{\gamma}}}=0 \nonumber \\
&b_{\underline{\underline{\alpha}}\beta}\tilde{a}_{\beta \underline{\gamma}}&=0, \; \;\;\;
b_{\underline{\underline{\alpha}}\beta}\tilde{b}_{\beta \underline{\underline{\gamma}}}=\delta_{\underline{\underline{\alpha}}
\underline{\underline{\gamma}}}  \label{w31} \\
&\tilde{a}_{\alpha \underline{\gamma}}a_{\underline{\gamma}\beta}&+\;\;\tilde{b}_{\beta \underline{\underline{\gamma}}}
b_{\underline{\underline{\gamma}}\beta}=\delta_{\alpha \beta} \nonumber
\end{eqnarray}
and define new variables which extend $\pi_{\alpha}\rightarrow (\pi_{\underline{\alpha}},
\pi_{\underline{\underline{\alpha}}})$\ to the full canonical transformation
\begin{eqnarray}
A_{\underline{\alpha}}\equiv \tilde{a}_{\beta \underline{\alpha}}A_{\beta}, \;\; 
A_{\underline{\underline{\alpha}}}= \tilde{b}_{\beta \underline{\underline{\alpha}}}A_{\beta}; \label{w32}
\end{eqnarray}
the inverse transformations read
\begin{eqnarray}
A_{\alpha}=a_{\underline{\beta} \alpha}A_{\underline{\beta}}+b_{\underline{\underline{\beta}} \alpha}
A_{\underline{\underline{\beta}}} \label{w33} \\
\pi_{\alpha}=\tilde{a}_{\alpha \underline{\beta}}\pi_{\underline{\beta}}+\tilde{b}_{\alpha 
\underline{\underline{\beta}}}\pi_{\underline{\underline{\beta}}} \label{w34}
\end{eqnarray}
We supply (\ref{w33}) with the analogous transformation for $u_{\alpha}$\
\begin{eqnarray}
u_{\alpha}=a_{\underline{\beta} \alpha}u_{\underline{\beta}}+b_{\underline{\underline{\beta}}\alpha}
u_{\underline{\underline{\beta}}} \label{w35} 
\end{eqnarray}
Inserting (\ref{w33}) $\div$\ (\ref{w35}) into the constraints $\pi_{\alpha}\approx 0, \; \chi_{\alpha}
\approx0, \; \tilde{\chi}_{\alpha}\approx0 $\ we conclude that these constraints are equivalent to the
following ones
\begin{equation}
\left.
\begin{array}{ccl}
\pi_{\underline{\alpha}}=0 \;-\; \hbox{primary} \\
\chi_{\underline{\alpha}} \equiv  a_{\underline{\alpha}\beta}(f_{\beta \gamma \delta}p_{\gamma}q_{\delta}
-v_{\beta})=0 \;-\;\hbox{secondary} 
\end{array}
\right\}  \;\;\hbox{I-st  class} \;\; \label{w36}
\end{equation}\\
\begin{equation}
\left.
\begin{array}{ccl}
\pi_{\underline{\underline{\alpha}}} = 0 \;-\;\hbox{primary}     \\
\chi_{\underline{\underline{\alpha}}} \equiv b_{\underline{\underline{\alpha}}\beta}
(f_{\beta \gamma \delta}p_{\gamma}q_{\delta}-v_{\beta})=0\;-\;\hbox{secondary} \\
\tilde{\chi}_{\underline{\underline{\alpha}}} \equiv A_{\underline{\underline{\alpha}}}=0\;-\;\hbox{secondary} 
\end{array}
\right\} \; \hbox{II-nd class} \label{w37}
\end{equation}\\
As always the Lagrange multipliers related to second - class constraints are fixed; here 
$u_{\underline{\underline{\alpha}}}=0.$\ On the other hand the arbitrary functions entering 
the general solution to Lagrange equations are $A_{\underline{\alpha}}(t);$\ consequently,
 $\dot{u}_{\underline{\alpha}}\equiv A_{\underline{\alpha}}(t)$\ are also arbitrary.\\
In order to put the theory in Dirac form let us cosider first the second - class constraints. The matrix of Poisson brackets
 of these constraints takes the form 
\begin{equation}
\left[
\begin{array}{ccl}
\{\pi_{\underline{\underline{\alpha}}},A_{\underline{\underline{\beta}}} \}&& \;\; \;\;\;\;\;0 \\
0 &&\;\; \{\chi_{\underline{\underline{\alpha}}},\chi_{\underline{\underline{\beta}}}\}
\end{array}
\right] \; \label{w38}
\end{equation}\\
where, on the constrained submanifold
\begin{eqnarray}
\{\chi_{\underline{\underline{\alpha}}},\;\chi_{\underline{\underline{\beta}}}\}=-
b_{\underline{\underline{\alpha}}\gamma}b_{\underline{\underline{\beta}}\delta}
f_{\gamma \delta \varrho}v_{\varrho}\equiv V_{\underline{\underline{\alpha}} 
\underline{\underline{\beta}}} \label{w39}
\end{eqnarray}
It is not difficult to prove from the definition of $b_{\underline{\underline{\alpha}}\beta}$\
that this matrix is nonsingular which confirms our classification (\ref{w36}), (\ref{w37}). One can now 
write out the basic Dirac brackets. First we note that due to the form of (\ref{w38}) the variables 
$A_{\underline{\underline{\alpha}}}$\, $\pi_{\underline{\underline{\alpha}}}$\ do not enter
the Dirac bracket - they disappear from the theory altogether. For the remaining variables we find 
\begin{eqnarray}
\{ A_{\underline{\alpha}}, \;\pi_{\underline{\beta}} \}_{D} &=& \delta_{\underline{\alpha} \underline{\beta}} \nonumber \\
\{q_{\alpha},\; q_{\beta}\}_{D} &=& \Delta_{\alpha \beta \xi \chi}q_{\xi}q_{\chi} \nonumber \\
\{p_{\alpha},\;p_{\beta}\}_{D} &=& \Delta_{\alpha \beta \xi \chi}p_{\xi}p_{\chi} \label{w40} \\
\{q_{\alpha},\;p_{\beta}\}_{D} &=& \delta_{\alpha \beta}\;+\;\Delta_{\alpha \beta \xi \chi}q_{\xi}p_{\chi} \nonumber
\end{eqnarray}
where
$\Delta_{\alpha \beta \xi \chi} \equiv v^{-1}_{\underline{\underline{\varrho}} \underline{\underline{\sigma}}}
b_{\underline{\underline{\varrho}} \gamma}b_{\underline{\underline{\sigma}} \tau}
f_{\gamma \alpha \xi}f_{\tau \beta \chi}$\ \\
It is not difficult to show that this theory is equivalent to the hamiltonian reduction for CM model. Let $v=\;(1,...,1)$\ be
the vector in defining representation of SU(N) and
\begin{eqnarray}
v_{\alpha}\equiv \nu\; Tr(\lambda_{\alpha}(v \otimes v^{+} - 1)),\label{w41} 
\end{eqnarray}
where $ \{\lambda_{\alpha}\}$\ is a basis of sU(N) such that $ [\lambda_{\alpha}, \;\lambda_{\beta}]
\;=\;if_{\alpha \beta \gamma}\lambda_{\gamma}, \;\;
 Tr(\lambda_{\alpha}\lambda_{\beta})\;=\;\delta_{\alpha \beta}.$\ 
Therefore, $G_{v}\subset SU(N)$\ consits of all matrices such that $Uv=e^{i \theta}v,\;i.e.$\
$ G_{v}\;=\; S(U(N-1) \times U(1)),$\ $dimG_{v}=(N-1)^{2}.$\ 
The initial set of dynamical variables (i.e. the "large" phase space to be reduced) consist of
$4(N^{2}-1)$\ variables $q_{\alpha}, p_{\alpha}, A_{\alpha}, \Pi_{\alpha}.$\ Now, the I-st 
and II-nd class constraints  are given by eqs.(\ref{w35}) and (\ref{w36}), respectively.
The primary constraints $\Pi_{\alpha}\approx 0$\ eliminate $\Pi_{\alpha}$\ altogether. The
secondary constraints $\chi_{\underline{\alpha}}\approx 0,\;\chi_{\underline{\underline{\alpha}}}
\approx 0$\ determine the preimage of momentum map. The remaining II-nd class secondary constraints 
$A_{\underline{\underline{\alpha}}} \approx 0$\ reduce the number of gauge fields to that of the dimension of residual 
gauge symmetry $G_{v}.$\ It is important to note that we are considering only the $G_{v}$-invariant sector
of our theory. This is because, on Hamiltonian picture level, the reduced phase space is \underline{not} the preimage of the momentum map but rather
the latter divided by the action of $G_{v}$\ group; therefore, all points on a given orbit are identified.
Any function on phase space, when reduced to the submanifold of constraints, depends a priori on 
$q_{\alpha}, p_{\alpha}$\ and $A_{\underline{\alpha}}.$\ The invariance under time-dependent $G_{v}$\ transformations implies that 
it does not depend on $A_{\underline{\alpha}}$\ and is a function of $G_{v}$-invariants built out of
$q_{\alpha}$\ and $p_{\alpha}.$\ To get a number of degrees of freedom of find theory let us note that the number
of first-class constraints (\ref{w35}) equals twice the dimension of $G_{v},\;2(N-1)^{2}.$\
The number of second-class ones (\ref{w36}) is $3(dimG-dimG_{v})=6N-6.$\ Therefore, there remains 
$4(N^{2}-1)-2(N-1)^{2}-(6N-6)=2N^{2}-2N$\ degrees of freedom which
is just the dimension of preimage of momentum map $N^{2}-1,$\ plus the number of $A_{\underline{\alpha}}$\
variables, $(N-1)^{2}.$\ For $G_{v}$-invariant functions the number of independent variables is
$2N^{2}-2N$\ minus the number of $A_{\underline{\alpha}}$\ variables, $(N-1)^{2},$\ minus the 
dimension of $G_{v},$\ again $(N-1)^{2},$\ which gives $2N-2,$\ as it should be.
Summarizing, the second-class costraints reduce the system to the one based on $G_{v}$\ invariance with 
"matter fields" $q_{\alpha},p_{\alpha}$\ being reduced to the 
preimage
of 
momentum
map.\\
In order to put the dynamics in the form which is explicitly equivalent to CM one
we can fix the temporal gauge $A_{\underline{\alpha}}=0.$\ 
Then all primary constraints become second - class and the modified Dirac bracket does not contain
$A_{\alpha}$\ and $\pi_{\alpha}$\ variables. The only constraints are now 
\begin{eqnarray}
\chi_{\underline{\alpha}} \equiv a_{\underline{\alpha} \beta}(f_{\beta \gamma \delta}
p_{\gamma}q_{\delta}-v_{\beta})=0\; \hbox{ -  I-st class} \nonumber \\
\nonumber \\
\chi_{\underline{\underline{\alpha}}} \equiv b_{\underline{\underline{\alpha}} \beta}
(f_{\beta \gamma \delta}p_{\gamma}q_{\delta}-v_{\beta})=0\; \hbox{ - II-nd class} \label{w42} \\ 
\nonumber 
\end{eqnarray}
which is equivalent to
\begin{eqnarray}
f_{\beta \gamma \delta }p_{\gamma}q_{\delta}-v_{\beta}=0 \label{w43}
\end{eqnarray}
while hamiltonian equations become
\begin{eqnarray}
\dot q_{\alpha}&=&p_{\alpha} \label{w44} \\
\dot p_{\alpha}&=&0 \nonumber
\end{eqnarray}
Identifying $A\equiv q_{\alpha}\lambda_{\alpha} \;,\;\; B\equiv p_{\alpha}\lambda_{\alpha}$\
we get \\
\begin{eqnarray}
&&\omega=Tr(dB \wedge dA), \;\;\;H=\frac{1}{2} TrB^{2} \nonumber 
\end{eqnarray}
\begin{equation}
\left. \nonumber
\begin{array}{ccl}
&&\dot A=B \;\;\;\;\;\;\;\;\;\;\;\;\;  \\
&&\dot B=0  \\
\end{array}
\right\} \;\;\;\;\label{w45}
\end{equation}
and the standard constraint 
\begin{eqnarray}
 [A,B]=i\nu(v \otimes v^{+}-I) \label{w46}
\end{eqnarray}
the  invariant under time independent $G_{v}-\hbox{transformations}$\
 solutions to this constraint are provided by C-M model
as it was described in introduction.

\end{document}